# Structural, Dynamic, and Vibrational Properties during Heat Transfer in Si/Ge Superlattices: A Car-Parrinello Molecular Dynamics Study

Pengfei Ji and Yuwen Zhang[1]

Department of Mechanical and Aerospace Engineering, University of Missouri, Columbia, MO 65211, USA

Mo Yang

College of Energy and Power Engineering, University of Shanghai for Science and Technology, Shanghai 200093, China

## Abstract

The structural, dynamic, and vibrational properties during the heat transfer process in Si/Ge superlattices, are studied by analyzing the trajectories generated by the ab initio Car-Parrinello molecular dynamics simulation. The radial distribution functions and mean square displacements are calculated and further discussions are made to explain and probe the structural changes relating to the heat transfer phenomenon. Furthermore, the vibrational density of states of the two layers (Si/Ge) are computed and plotted to analyze the contributions of phonons with different frequencies to the heat conduction. Coherent heat conduction of the low frequency phonons is found and their contributions to facilitate heat transfer are confirmed. The Car-Parrinello molecular dynamics simulation outputs in the work show reasonable thermophysical results of the thermal energy transport process and shed light on the potential applications of treating the heat transfer in the superlattices of semiconductor materials from a quantum mechanical molecular dynamics simulation perspective.

---

[1] Corresponding author. Email: zhangyu@missouri.edu

# Nomenclature

$E$      atomic-scale energy, Ha

$M$      atomic mass, a. u. = 0.00054857990943 a. m. u

$\boldsymbol{p}$      polarization vector

$\boldsymbol{q}$      wave vector

$\boldsymbol{R}$      atomic position vector, Bohr

$T_s$      kinetic energy of a non-interacting system, Ha

$U$      potential, Ha

$\nu$      vibrational frequency of atom

$V_{ext}$      external potential, Ha

Greek

$\alpha, \beta$      atomic species

$\varepsilon^{KS}$      extended Kohn-Sham energy function, Ha

$\varphi$      electronic one particle orbital

$\delta_{ij}$      orthonormality relation, $\langle \varphi_i | \varphi_j \rangle$

$\mu_i$      fictitious mass, a. u.

$\mathcal{L}$      Lagrangian form of energy, Ha

$\Psi$      electronic wave function at ground state

| | |
|---|---|
| $\Lambda_{ij}$ | Lagrange multiplier imposing on the orthonormality constraints |

*Subscripts and Superscripts*

| | |
|---|---|
| $CP$ | Car-Parrinello |
| $N$ | union of all the positions |
| $c$ | correlation energy in DFT |
| $H$ | Hartee energy in DFT |
| $x$ | exchange energy in DFT |
| $e$ | electron |

# 1. Introduction

Superlattices were discovered in the early 20th century through their special X-ray diffraction patterns. For semiconductor superlattices, the idea was originated by Esaki and Tsu [1]. As envisioned, there are two kinds of superlattices: one is compositional and the other is doping. The compositional superlattices are consisted of alternating super-thin layers of two more other kinds of materials that have nearly the same lattice parameters. In general, the thicknesses of individual layer are in the range between angstroms to nanometers. In the compositional superlattices, the electronic potential is modulated on a length scale that is shorter than the electron mean free path by the layer composition. As for the doping superlattices, they cover alternating n- and p-type layers of a single semiconductor. The electric fields generated by the charged dopants play the role of modulating the electronic potential. The cutting-edge nanotechnology has been the driving force to lead the development of novel applications and devices based on the superlattices. One of the most important parts of the design of superlattices is the thermal management. The microsystem will fail or even break due to overheat, if the thermal energy generated inside the superlattices are not transported away efficiently. Understanding the heat transfer process occurring in the semiconductor superlattices has been an issue of great importance in the composited semiconductors fabrication. The interfacial thermal resistance plays a critical role in the heat conduction process between the neighboring layers. In this work, heat transfer process for the superlattices composed of silicon and germanium layers with periodicity will be investigated.

With the reduction of the size of structure, the thermophysical transport properties of nanostructures are significantly different from those of the bulk values of the constituent materials. They are no longer intrinsic parameters as in macroscopic region. As demonstrated, the thermal conductivity of silicon renders a tendency of decrease when the thickness of a layer is less than 1000 Å [2]. Hence, the size effect should receive enough attention in practical engineering applications, such as the thermoelectric energy conversion and the microelectronic thermal management. During the past two decades, in the fields of fabrication of microelectronic and optoelectronic devices, thermoelectric and thermionic devices [3], the heat transfer phenomenon in superlattices is increasingly recognized and drawn considerable attentions [4]. When it comes to investigate the energy transport efficiency, usually, it refers two opposite goals, namely, to enhance or decrease the heat transfer inside the superlattices materials. For the case of microelectronics devices, which are composited with layers of different conductive materials, the object is to increase the thermal conductance between the neighboring layers in order to move the heat generated by Joule effect and realize the purpose of efficient cooling. On the contrary, as for the thermoelectric materials, to lower the overall thermal transport efficiency by mixing dissimilar crystalline materials and artificially adding grain boundaries and gaps is the primarily goal. From the nanoscopic point of view, the thermal energy is transported by different energy carriers like phonons, electrons, or molecules. For semiconductor superlattices, phonons are the major energy carriers that move through the stack of thin layers. In a material with many layers, such scattering effects would accumulate so that the thermophysical properties of semiconductor superlattices are strongly impacted by phonon motion inside the layers and scattering at the layer interfaces. It has been a high demand to understand the atomic level mechanism during the heat transfer process in superlattices.

The microstructure changes of the superlattices induced by the heat flow penetrating through the alternating layers will also affect the thermal transporting efficiency. As described in experimental studies on Si/Ge superlattices [5], the nanostructure based on samples grown by metal-organic chemical vapor deposition on GaAs substrates with Ge buffers, showed reduced thermal conductivity compared with the bulk materials. In order to come up with appropriate explanations for the experimental findings, researchers proposed various models for the solution of heat conduction in superlattices, e.g., the Boltzmann transport equation [6], the phonon group velocity reduction due to zone folding [7,8], and the atomistic Green's function methods [9,10]. Moreover, some researchers even suggested the potential application of control the conduction of heat by mass-loading nanotubes with heavy molecules [11], manipulating the band formation [12] and creating phonon-surface roughness heat transfer condition [13].

Besides the above theoretical modeling and analysis, molecular dynamics (MD) simulation paves a new way to investigate the thermal energy transport phenomenon in superlattices. Since the MD method was firstly introduced by Alder and Wainwright in the 1950s [14,15], MD has been widely applied to study the structural, dynamical, and thermophysical properties of complex systems. The MD simulation results can be used to determine macroscopic thermodynamic properties of the system based on ergodic hypothesis [16], in which the statistical ensemble averages are assumed to be equal to time averages of the system. It has been termed "statistical mechanics by numbers" of predicting the future by animating the natural forces and enable us sight into the thermal energy transport in atomic scale. In terms of classical MD, the interatomic force and acceleration used for generating trajectories are based on the first order and second order derivatives of predefined empirical potential functions. The classical MD calculation of thermal interface conductance was carried out by Chalopin *et al.* [17] using equilibrium MD with variable thickness, and showed the evanescent modes when the layer thickness is below 5 $nm$. For nonequilibrium MD simulations, a lattice thermal conductivity for GaAs/AlAs was calculated by Imamura *et al.* [18] with hot and cold reservoirs to produce heat flow running from one end to the other.

Ji and Zhang carried out ab initio simulations spanning from heat conductions to thermal radiation [19]. For the cases of rough interfaces, Termentzidis *et al.* [20] obtained the in-plane thermal conductivities in binary Lennard-Jones superlattices by setting up different characteristic heights of roughness.

However, due to the limitation of available potentials and the transferability problems, classical MD fails to yield accurate results for some cases. When the transport process involves the motion of electrons, the classical MD faces another shortcoming to integrate in the electronic degrees of freedom. Fortunately, the ab initio MD overcomes the aforementioned drawbacks of classical MD to get a more accurate representation of force field and to obtain the electronic behavior by using the density functional methods. The challenge to calculate the full set of quantum mechanical many body forces for each instantaneous atomic configuration of the system can be solved by the Car-Parrinello (CP) [21] MD simulation, which simplifies quantum mechanical adiabatic time scale separation of fast electronic and slow nuclear motions by transforming that separation into a classical mechanical adiabatic energy scale separation in the framework of dynamics theory. CPMD simulation is aimed to perform a simulation in which the interatomic forces are derived directly from the electronic ground state using the Hellman-Feynman theorem. In CPMD, for a given atomic configuration, the Born-Oppenheimer (BO) [22] potential energy surface is obtained by minimizing the total energy functional with respect to one electron states.

In this work, our main objective is to elucidate the nonequilibrium characteristics of the microstructure, translation dynamics and vibrational density of states of Si/Ge superlattices during heat transfer process. We adopt the ab initio MD method to simulate the heat conduction occurring in the cross-plane direction of symmetrically strained Si/Ge supperlattices. This paper is organized as follows: In Section 2, the fundamental theory and algorithm of CPMD, as well as the analysis approaches of radial distribution function (RDF) calculations mean square displacements (MSD) and vibrational density of states (VDOS) are described in detail. In Section 3, the modeling details of the periodic Si thin layer and Ge thin layer on the superlattices, as well as the setup for the initial non-equilibrium state between the neighboring layers and the subsequent NVE ensemble simulation are given. In Section 4, the calculated temperature curves corresponding to the evolution of simulation process is plotted and further comparisons of the RDF and MSD curves of the Si/Ge layers at specified moments of the equilibration time are made. Meanwhile, analyses and discussions for the differences among the different figures are carried out to show the influences of heat transfer on the structures and dynamics of the Si/Ge superlattices. Moreover, the vibrational density of states (VDOS) of the two layers are plotted and analyzed to investigate the contribution of different frequency phonons to the thermal transport efficiency.

## 2. Theory and Methods

### 2.1 Ab Initio Molecular Dynamics

We perform ab initio MD simulations using Car-Parrinello scheme [21], which uses the density functional theory (DFT) to derive the interatomic force constants. The CPMD is chosen not only because it bears the advantage of being free from the empirical parameters and its general applicability, but also it cuts the computational expenses of BOMD by simplifying the electrons in a single state. The significant difference that distinguishes the CPMD from BOMD is the introduction of an artificial fictitious dynamics for electronic degrees of freedom. For the CPMD simulation, in order to transform the quantum adiabatic time-scale separation of fast electronic and slow nuclear motion into classical adiabatic energy-scale separation for the dynamics, the fictitious dynamics for electronic degrees of freedom is introduced. The CPMD method efficiently decouples the nuclear time evolution and electronic minimization via an implicit adiabatic dynamics approach. The introduced fictitious dynamics for the electronic orbitals provides initially minimum electronic orbitals for an initial nuclear configuration, and enables the nuclear motion adiabatically, which keeps the approximately minimized configuration at each step of the molecular dynamics simulation. Thus, the introduced fictitious electronic dynamics ensures that the electronic wavefunction keep automatically minimized to its ground state along with the propagation of nuclei. Furthermore, comparing with BOMD that needs to solve the electronic wavefunction self-consistently at each MD step, the computational load of CPMD is greatly reduced.

The CPMD method has already been described elsewhere [21,23,24] in detail so that only fundamentals that is relevant to this work will be presented here. It should be pointed out that there is no additional artificial nonequilibrium force introduced due to the fictitious dynamics of electrons. To maintain the adiabatic separation between the fictitious electron dynamics and the nuclear dynamics, the thermostats (Nose-Hoover method in this work) are imposed on the electronic subsystem. It avoids the Schrodinger equation to be solved at each step in the simulation. During the CPMD simulation process, the electronic temperature are kept at a much lower fictitious value comparing with the nuclear temperature, so that the electronic subsystem is close to its instantaneous minimum energy. According to the Kohn-Sham (KS) theory, the difficult many-body system can be replaced with an independent particular system, which leads to the ground state energy of the interacting system of electrons with classical nuclei at position $\{r_{n,I}\}$. The minimized KS energy [25] can be expressed as:

$$E_{min}^{KS} = T_s + \int V_{ext}\rho(r)\,dr + E_H + E_x + E_c \tag{1}$$

where $T_s$, $\int V_{ext}\rho(r)\,dr$, $E_H$, $V_x$ and $V_c$ denote the kinetic energy, the electron-ion energy, the classical Hartree energy, the exchange energy and the correlation energy, respectively.

The extended KS energy $E^{KS} - \sum_{ij}\Lambda_{ij}(\langle\varphi_i|\varphi_j\rangle - \delta_{ij})$ is denoted as $\varepsilon^{KS}$ for simplicity. In classical mechanics, the force on nuclei is obtained from the derivate of Lagrangian with respect to nuclear positions. Similarly, by considering the extended KS energy function $\varepsilon^{KS}$ to be dependent on $\boldsymbol{R}^N$ and $\boldsymbol{\varphi}_i$, we can construct a Lagrangian functional derivative with respect to the orbitals, which can also be interpreted as classical force fields:

$$\mathcal{L}_{CP}[\boldsymbol{R}^N, \dot{\boldsymbol{R}}^N, \{\boldsymbol{\varphi}_i\}, \{\dot{\boldsymbol{\varphi}}_i\}] = \sum_I \frac{1}{2}M_I\dot{\boldsymbol{R}}_I^2 + \sum_i \frac{1}{2}\mu_i\langle\dot{\boldsymbol{\varphi}}_i|\dot{\boldsymbol{\varphi}}_i\rangle - \varepsilon^{KS}[\{\boldsymbol{\varphi}_i\}, \boldsymbol{R}^N] \tag{2}$$

Therefore, the corresponding Car-Parrinello equations of motion are obtained from the associated Euler-Lagrange equations:

$$M_I \ddot{R}_I(t) = -\frac{\partial \varepsilon^{KS}}{\partial R_I} = -\frac{\partial \left[E^{KS} - \Sigma_{ij}(\langle \varphi_i | \varphi_j \rangle - \delta_{ij})\right]}{\partial R_I} = -\frac{\partial E^{KS}}{\partial R_I} + \Sigma_{ij} \Lambda_{ij} \frac{\partial \langle \varphi_i | \varphi_j \rangle}{\partial R_I} \quad (3)$$

$$\mu_i \ddot{\varphi}_i(t) = -\frac{\delta \varepsilon^{KS}}{\delta \langle \varphi_i |} = -\frac{\delta \left[E^{KS} - \Sigma_{ij}(\langle \varphi_i | \varphi_j \rangle - \delta_{ij})\right]}{\delta \langle \varphi_i |} = -\frac{\delta E^{KS}}{\delta \langle \varphi_i |} + \Sigma_j \Lambda_{ij} |\varphi_j\rangle \quad (4)$$

According to Eqs. (3) and (4), the idea of getting the time scale separation of fast electronic and slow nuclear motion is realized by transforming that into classical mechanical adiabatic energy scale separation in the framework of DFT. Therefore, the simultaneous time evolution of both the ionic and the electronic degrees of freedom can be determined by integrating Eq. (3) and (4).

## 2.2 Radial Distribution Function

The radial distribution function $g(r)$ in a system composed of atoms describes how density varies as a function of distance from a reference particle. It is also referred to as pair distribution function or pair correlation function, which describes how, on average, the atoms in a system are radially packed around each other. We can get the probability to find an atom in a shell $dr$ at the distance $r$ of another atom chosen as a reference point from the calculated RDF.

As shown in Fig. 1, the number of atoms inside the shell region between $r$ and r+$dr$ for a given system can be expressed in term of RDF $g(r)$ as:

$$dn(r) = \frac{N}{V} g(r) 4\pi r^2 dr \quad (5)$$

where $N$ and V represent the total number of atoms in the model and its volume, respectively. For a system that contains more than one kind of chemical species, the RDF is referred to as partial radial distribution function, which means the density probability for an atom of the $\alpha$ species to have a neighbor of the $\beta$ sepcifices at given radial distance $r$. It can be computed according to the following equation:

$$g_{\alpha\beta}(r) = \frac{dn_{\alpha\beta}(r)}{4\pi r^2 dr \rho_\alpha} \tag{6}$$

where $\rho_\alpha = \frac{N_\alpha}{V} = \frac{Nc_\alpha}{V}$ and $c_\alpha$ is the concentration of atomic species $\alpha$.

In addition, the RDF can be used in conjunction with the interatomic pair potential function to calculate the internal energy of the system.

## 2.3 Mean Square Displacement

Mean square displacement is a parameter of measuring the spatial extent of random motion. In processing the MSD, the square distances in $x$, $y$ and $z$ are calculated and then average the square distances, taken over all atoms. The MSD is mathematically defined by the relation:

$$MSD(t) = \langle r^2(t) \rangle = \langle |r_i(t) - r_i(0)|^2 \rangle \tag{7}$$

where $r_i(t)$ represents the position of the atom $i$ at the time $t$ and $\langle \cdots \rangle$ is denoted as an average on the time steps and the particles. The MSD is a significant quantity which relates to the information on the diffusion of atoms. In general, the MSD for atoms in liquid or gas grows linearly in time. However, if the system is solid, then MSD is "saturated" and the kinetic energy is not sufficient to reach a diffusive behavior. At short times, the plot of MSD and time is not linear. In the model of our simulation, the simulation time is a very short period spanning from femtosecond to picosecond. Further more details and discussions of the MSD plots obtained our simulation results will be given in Section 4.

## 2.4 Vibrational Density of States

In condensed matter physics, the vibrational density of states of a system is used to characterize the number of states per interval of energy at each energy level that are available to be occupied by phonons/electrons. It is an important feature of a solid system [26,27], on which the thermodynamical properties depend. At absolute zero temperature, a crystal lattice contains no phonons, which is named as ground state. At non-zero temperature, a crystal lattice has an energy that is not constant, but fluctuates randomly around the mean value.

The calculation of vibrational properties is relied on the velocity autocorrelation function (VAF). By performing Fourier transform of the VAF averaged over all atoms, the normalized VDOS of $N$ atoms system is given by

$$g(v) = \int_{-\infty}^{+\infty} \frac{\sum_{i=1}^{N} \langle \mathbf{V}_i(t) | \mathbf{V}_i(0) \rangle}{\sum_{i=1}^{N} \langle \mathbf{V}_i(0) | \mathbf{V}_i(0) \rangle} e^{2\pi i v t} dt \tag{8}$$

where $\mathbf{V}_i$ denotes the velocity of the $i$th atom.

The vibrational modes can be described by phonon-like wave packets, which appear in VDOS as peaks. The maxima of peaks can be used to determine the dominant frequency of the mode. We can judge what extent the phonon picture applies to the heat transfer system by comparing the clear and isolated peaks in the VDOS plot.

## 3. Modeling and Simulation

The ab initio MD calculations were carried out based on the plane-wave pseudopotential implementation of density functional theory, which is written in the latest CPMD package version 3.15.3 [28]. We established a supercell that contains atoms $4(width) \times 4(length)$ as the cross section basis and 8 atoms equally assigned for each layer (as is shown in Fig.2 by VMD [29]). For the sub-model of silicon layer, we set the lattice constant as 5.431 Å [30] in the Cartesian coordinate system, while for the sub-model of germanium layer, the lattice constant was 5.568 Å [31]. Both the silicon and germanium have zinc-blende structure. Because the parameters of the supercell are set 11.316Å for the length of the cross-section square and 22.632 Å in the height, the neighboring Si/Ge layers can be regarded as closely attached to each other. Furthermore, the same kind of lattice structure makes the interface match well. The lattice structure of Si and Ge are both diamond cubic and the two layers are in direct contact (without any gap separating the layers). Therefore, we can treat the heat transfer process as heat conduction, rather than radiation. To simulate the alternating super-thin layers of silicon and germanium, we employed periodic boundary conditions in all three directions during the entire simulation process. A schematic view of the Si/Ge superlattices is shown in Fig. 3.

To start the CPMD simulation, wavefunction optimization of the system was the first step and further MD simulations were based on the optimized system. If a MD simulation is executed from a non-converged wavefunction, in the consequent work, the electrons maybe far away from the Born-Oppenheimer surface and the results produced will be unphysical. In our simulation, the CPMD calculations were executed based on the LDA of the exchange-correlation functional augmented by Becke–Lee–Yang–Parr (BLYP) [32,33] generalized gradient corrections. The valence electrons were treated explicitly and electron-ion interactions were described by norm-conserving pseudopotentials generated. We chose the Trouiller-Martins [34] norm-conserving nonlocal pseudopotentials to represent the core electrons for both silicon and germanium and the maximum $l-$quantum numbers were set as P. The Kohn-Sham orbitals had the periodicity of the supercell and were expanded in plane waves up to a kinetic energy cutoff of 30 Ry. The $\Gamma$ point $[k=(0,0,0)]$ was used to sample the Brillouin zone of MD supercell. After obtaining the optimized wavefunciton, the electronic degrees of freedom were quenched to the Born Oppenheimer surface. We started the CPMD simulation with NVT ensemble adopted to form the steady non-equilibrium state with a time step of $4.0\ a.u.(about\ 0.096,755\ fs)$ to integrate the equations of motion, which was found to produce a satisfactory convergence in the wavefunction optimization and geometry optimization. To keep the constant temperatures of the layers, silicon $800\ K$ and germanium $300\ K$, the Nose-Hoover thermostats [35] were imposed on ions and electrons (with a fictitious electronic mass of $400\ a.u.$) for each degree of freedom. We ran the NVT simulation for 20,000 time steps to make sure that the layers had sufficient time to reach steady states, which meant that the temperatures differences are well established for the layers. Then, the thermostats were removed and the system was set adiabatically to launch the NVE simulation. It came to the most crucial point to keep the total energy of the two layers conserved and let the two sorts of thermal gradients evolve smaller and smaller spontaneously. The vibrations of atoms varied with the heat flow from silicon layer to germanium layer. During the entire CPMD simulation, atomic configurations and velocities were saved in every time step, then we extracted the trajectories at different time ranges and perform statistical computations of the RDF and MSD based on the program ISAACS [36]. In addition, the overall temperature curves of the two layers were plots with the evolution of simulation time. In the end, the normalized vibration spectra were also calculated as a supplementary part to probe the resonance mode of the two layers' thermal communication.

## 4. Results and Discussions

### 4.1 Temperature Evolutions with Simulation Time Steps

We calculated the temperatures of the two layers along with the proceeding of MD simulation and plotted the two temperature curves corresponding to the simulation time steps of the silicon and germanium layers, respectively. The temperatures are defined according to the kinetic theory by taking mean kinetic energy over all Si/Ge atoms in three directions. There are two purposes to monitor the temperature variations: (1) the temperature curves reflect general thermal states of the two layers and their difference, and (2) we can determine the moment to terminate the simulation at the point which the two temperature curves converges.

As is shown in Fig. 4, at the first $20,000$ time steps with the thermostats clapped on both layer, the temperate profiles oscillate around $800\,K$ for silicon layers and $300\,K$ for germanium layers. The oscillation magnitudes are inversely proportional to the square root of the numbers of atoms included in the calculation of the mean temperatures of the silicon and germanium subsystems. Since the mean temperature of silicon layer is higher than that of germanium layer, oscillation for silicon curve is reasonably more significant. After releasing the thermostats, heat conduction occurs due to the temperature gradient between the neighboring layers and the equilibration time is about $11.320\ ps$ ($117,000\ time\ steps$). In solid materials, the heat conduction process is interpreted as the vibrations of atoms coupling from each other that form waves of atomic displacements. Phonon is defined by the product of the vibrational frequency $v$ with the Planck constant $h$, which is the smallest quantized vibrational energy. The overall motions of atoms form the elastic waves that convey the thermal energy from one site to the other. As for semiconductors, due to the fact that the free electron density is much lower than that in metal, the contribution of electrons in thermal energy transport is neglected. By calculating under the framework of DFT theory, we can get accurate force constants acting on the atoms and then predict the motion of atoms. According to definition of phonon, one kind of spatially localized, quantized units of propagating vibration wave, one can associate the characteristic of phonons with the motions of atoms included in the trajectories obtained from MD simulation. As reported by Luo and Lloyd [37] equilibration time for silicon-germanium-layer ($750\ K$ versus $450\ K$) with the thickness of $5.658\ Å$ is $12.772\ ps$ ($132,000\ time\ steps$), which is close to and at the same order of magnitude with the simulation result $11.320\ ps$ in the current work.

## 4.2 Microstructures at Different Simulation Stages

In order to understand the structural property changes through the thermal energy transport process in detail, we studied the radial distribution function between the sites of different Si/Ge atoms. As an important structural characteristic of Si/Ge superlattices, we compute it based on the trajectories of atoms produced at different simulation stages from the initial nonequilibrium state to the final dynamical heat exchange moment. We extracted the trajectories of 5,000 time steps in the stage with thermostats added, every 5,000 time steps at the simulation moments of 25,000, 55,000, 85,000, 115,000 in the equilibration time stage, and the last 5,000 times steps before the end of simulation (the final equilibrium state). In the post-processing of our simulation result, we chose an atom (Si/Ge) in the system and drew around it a series of concentric spheres with different radii. At regular intervals a snapshot of the system is taken and the number of atoms found in each shell is counted and stored. The average number of atoms in each shell is calculated. For the object materials to be investigated is solid Si/Ge, in which the atoms vibrate at almost fixed positions rather than continually move from one place to the other. Hence, we chose the maximum distance in real space as the lattice constant of germanium atom with 5.658 Å. The plots of RDF are shown in Figs. 5-10. In our RDF calculations, we set the atomic radii (the mean distance from the nucleus to the boundary of the surrounding cloud of electrons) as 1.11 Å and 1.22 Å for silicon and germanium, respectively. At short separations, the RDF is zero. This indicates the effective width of the atoms, since they cannot be any closer. Usually, for liquids and gases, at high temperature, the peaks are broad, indicating thermal motion, while at low temperature, they are sharp. For the crystalline materials, where atoms are strongly confined in their positions, the peaks are particularly sharp. Due to the strong repulsive forces, at short distances, $g(r)$ is almost zero. Comparing Fig. 5 with the other five figures drawn since the stage of releasing the temperature control (see Figs. 6-10), we can find that the RDF curves for silicon and germanium in the former is more obviously separated than the latter five ones. In the constant temperature control stage of NVT simulation, there are no energy transfer and thermal communication between the two layers and the atomic vibrations solely depend on the Nose-Hoover algorithm. Thus, Fig. 5 reflects the structural properties of silicon ($800\ K$) and germanium ($300\ K$) at constant temperatrue. To make the calculated RDF have statistical sense, we take 5,000 time steps trajectories, even though temperatures of the two layers are dynamically varying due to internal energy change induced by thermal energy transport. As is shown in Figs. 6-10, the $g(r)[Ge, Ge]$ curve gradually approaches to the $g(r)[Si, Si]$ near the end of simulation. With the progress of heat transfer, $g(r)[Ge, Ge]$ starts to overlap with $g(r)[Si, Si]$ (see Fig. 8), which means the probabilities to find atoms at a certain position are approximately equal. Such kind of overlapping regions indicates the oscillation amplitudes of atoms for both silicon and germanium are equal and have the same probabilities. We can even find the two curves almost overlap each other near to the final thermal equilibrium stage (see Fig. 10).

## 4.3 Translation Dynamics at Different Simulation Stages

We calculated the overall MSD and the directional MSD for the two layers at different simulation stages at the same sampling points as those of RDA in Section 4.2.1. The MSD plots are illustrated in Figs. 11-16 with a sampling time range of $483.775\ fs$ (5,000 $time\ steps$). The profiles of the plotted MSD agree with the ones reported in [38]. It can be seen that the maximum MSDs for silicon and germanium layers are significantly different at the initial non-equilibrium stage (see Fig. 11). Since the temperature of the silicon layer ($800\ K$) is greater than that of the germanium layer ($300\ K$), the distance of motion at each time should be longer for silicon atoms. Since we set the dimensions for each layer with equal lengths and the layers are composed of $4 \times 4 \times 4$ atoms, the internal energy is equally distributed in the three dimensions. Comparing curves among the Figs. 11-16, we can conclude that the MSDs for germanium layer gradually become higher while the MSDs for silicon layer render a tendency of lower and lower. It indicates the atoms' displacements change agree well with the temperatures evolution. As it is shown in Fig. 16, at the final dynamic equilibrium stage with nearly equivalent amount of thermal energy absorption and emission, compared with curves in Fig. 15, the MSD values for silicon gets lower and germanium gets greater. But due to the mass difference of silicon and germanium atoms, the overall MSD of silicon should be greater than that of germanium.

## 4.4 Vibrational Density of States Analyses

Thermal properties, such as the heat capacity and thermal conductivity as well as some other kinds of material properties are strongly impacted by the vibrational density of states (VDOS). Hence, to get a profound grasp of the laws governing the vibrational properties of superlattices is of high technological and fundamental interest. To gain a further insight into the heat conduction process between the neighboring layers, we calculated the VDOS or the power spectra by performing the Fourier transformation of the velocity autocorrelation function. Figure 17 shows the distribution of normalized power density versus the vibration frequency (arbitrary unit, a.u.). A comparison of VDOS for experiment and computer simulation can be found in [39]. The distinguishable peaks of silicon and germanium corresponding to middle and high frequency vibrational modes are isolated distribute in the 3 a.u. and 5 a.u., respectively. The strong peaks lose height rapidly and the densities of distribution are relatively smaller than those in lower frequency region. From the vibration point of view, due to the mass difference, the maximum VDOS of Si (in high frequency region) is independent from that of Ge in high frequency region. Additionally, because of higher temperature of Si than that of Ge, the mainly isolated peak of Si is located at higher frequency regions. Nevertheless, there are still some VDOS overlapping areas existing in the in the frequency range of $0.5 - 2.5$ a.u., which demonstrates in thermal communication occurring between the neighboring layers inside the superlattices. For the phonons with discrete peaks in the frequency region above 2.5 a.u, as discussed in section 4.1, because of the difference of the materials, it will cause mismatch of phonon transport. Taking the superlattices as composite, phonon transport across the interface is incoherent heat transfer, which is the dominant factor of high thermal resistance. Luckyanova *et. al* [40] showed that the anharmonic and interface rates at different frequencies and the interfacial scattering of high frequency phonons led to a reduction in their heat carrying ability and directly induced the overall decrease in the thermal conductivity of superlattices. However, the phonon transport can also be treated as coherent if we regard the superlattices as a "homogeneous" material. For the phonon with overlap peaks below the frequency of 2.5 a.u, thermal diffusion occurs. In the Casimir classical size effect regime, broadband phonons are excited at one boundary and then traverse the internal region of the film ballistically before reaching the boundary of the layers [40]. Thus, phonon propagation is coherent inside the layer. When the vibrational phonons encounter the interface between the neighboring layers, the overlapping areas in the low frequency region of VDOS make it clear that such kind of vibrational phonons locating in low frequency region are the main contributors to facilitate the heat transfer. Thus, there is also a proportion of the phonon energy that can be coherently delivered to the neighboring layer. Namely, the interfacial heat transfer is coherent phonon heat conduction coexisting with incoherent phonon heat conduction. The proportion of coherent heat transfer has also already been discovered and investigated by Luckyanova *et al*. [40] with both experimental support and simulations verification. As explained in Ref. [40], under the constraint of fixed thickness of superlattice, the fraction of coherent phonon heat conduction to the overall heat conduction changes due to the temperature variations and periods in the SL. However, in the present work, the thickness and temperature of the Si/Ge superlattices are predefined, and the periodic boundary conditions are added in all x-, y- and z- directions. Therefore, the fractions of coherent phonon and incoherent heat conduction do not vary with the above three parameters. To interpret the mechanism in a physical point of view, as shown in the calculated VDOS in Fig. 17, the low frequency phonons correspond with long mean free paths and therefore enable the phonons propagate longer distance through the layer interfaces, which denotes the coherent heat conduction (see the overlapping spectrum in Fig. 17). Whereas the incoherent heat conduction is characterized by the middle and high frequency phonons (see the discrete spectrum in Fig. 17). Therefore, the coherent heat conduction coexists with incoherent heat conduction during the interfacial heat conduction process. The main contributors facilitating the heat transfer is the coherent phonons.

## 5. Conclusions

We present an ab initio molecular dynamics study of the microstructure, translation dynamics and vibrational density of states of Si/Ge superlattices. The simulation captures the essential features that the behavior of thermal energy transporting along the direction of the temperature gradient. We adopt non-equilibrium MD simulations starting from the initial temperature of $800\ K$ for silicon layer and $300\ K$ for germanium layer. The entire simulation process was conducted by monitoring the temperatures of the two layers and the simulation was finally terminated when the system reached thermal equilibrium state. The computed RDFs and MSDs of each layer at different simulation stages demonstrate significant changes of the superlattices structure and their relating characteristics. The structural changes are in consistent with the thermal motion of the atoms inside each layer. After obtaining the final equilibrium state between the layers, from the perspective of probability to find an atom at given place, RDF curves of the two layers show almost the same probabilities, which reflect the same kinds of average structural responses at thermal equilibrium. The comparisons of lateral and longitudinal results of MSDs at the same simulation moments and evolutional moments at different NVE simulation stages clearly show the MSD values variations as the heat conduction occurs. It provides us useful information to design superlattice materials when we would like to lower the heat transfer efficiencies across interfaces. Because the interface transmission is crucial for building a bridge to calculate thermal energy transport from pure materials to compositing superlattices, the computed VDOS from ab initial MD and the overlap areas of spectra indicate phonons locating in low frequency region indicate they are the main contributors to enhance the thermal energy transport efficiency through the interface of the thin layers. In addition, interface roughness is a crucial factor that impedes the heat transfer. But for the model we established, the compositing materials are with similar crystal structures and are well matched in the initial simulation step. Since the Si/Ge superlattices are often applied as thermoelectric material, to achieve the low thermal conductivity is of the great importance. Feasible approaches are modulating the vibrational frequencies to reduce the coherent heat transfer in the low frequencies regions and maximally increasing the phonon scattering at the interfaces and layer boundaries. We will introduce artificial patterns in the contacting surface of the layers to probe the heat transfer process and its efficiency in our further work.

Our work demonstrates the great potential of Car-Parrinello MD simulation of nanostructure materials. With the advantages over classical MD simulation and other numerical modeling methodologies, it is also possible to broaden the Car-Parrinello MD simulation to heat transfer problems involving the participation of electron interactions and transport in the future.

## Acknowledgment

Support for this work by the U.S. National Science Foundation under grant number CBET- 1066917 and National Natural Science Foundation of China under grant number 51129602 is gratefully acknowledged.

**Figure Captions**

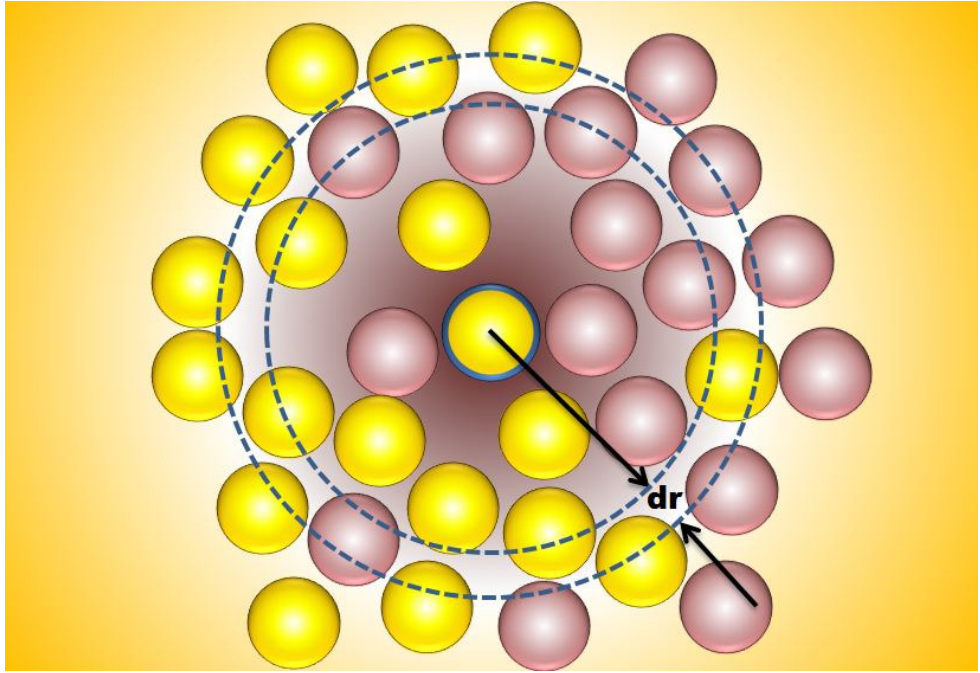

**Fig. 1**

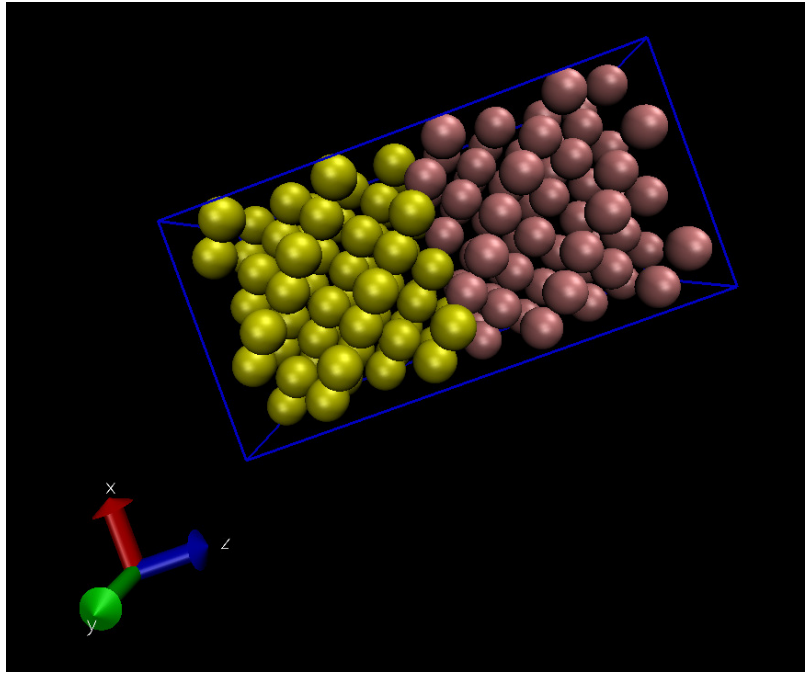

**Fig. 2**

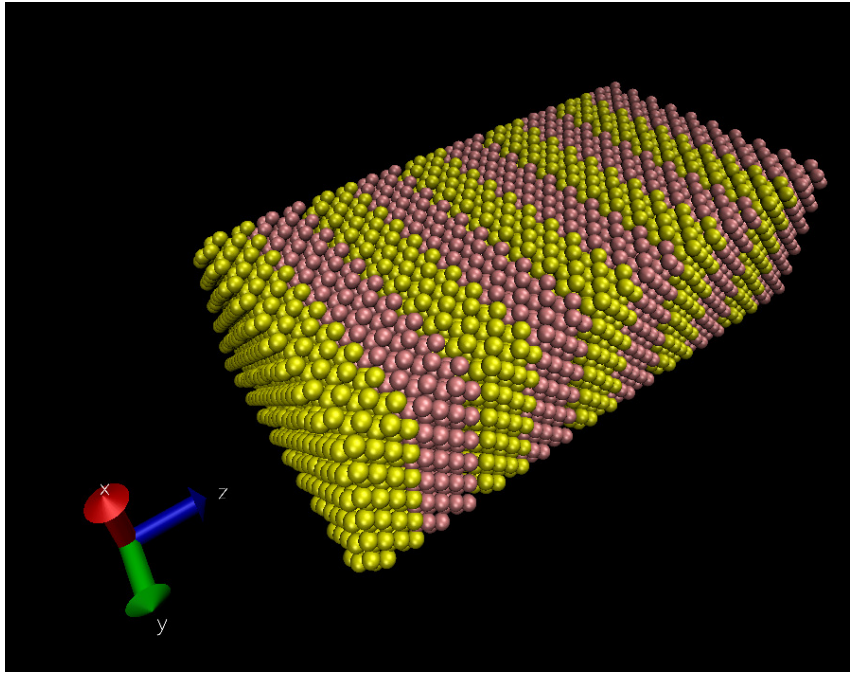

**Fig. 3**

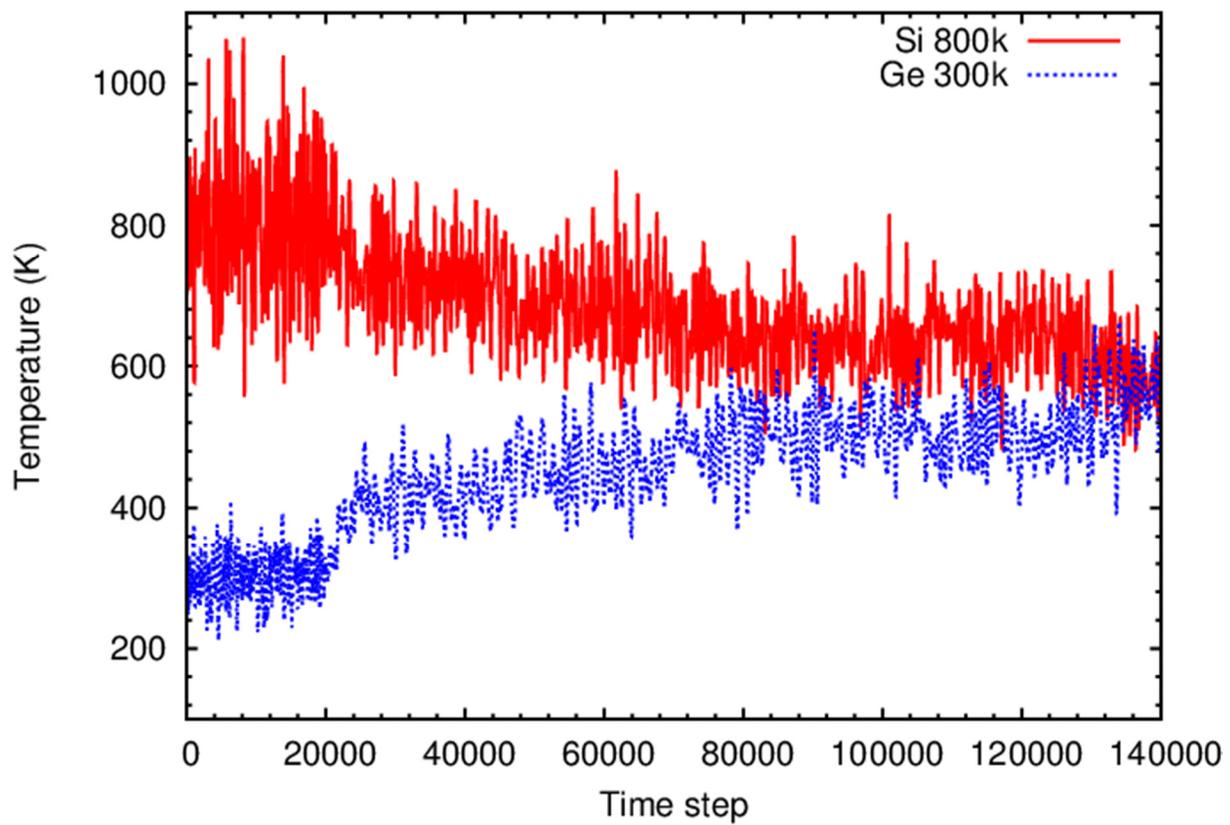

**Fig. 4**

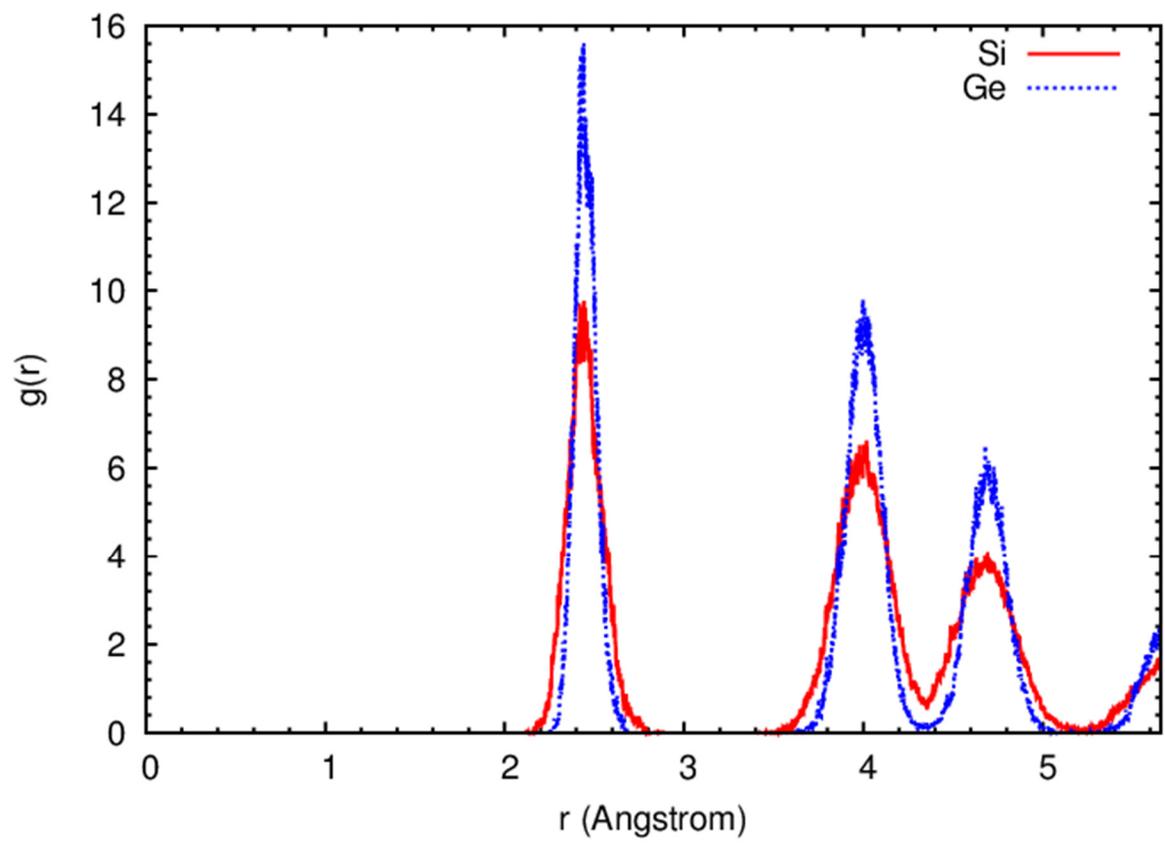

**Fig. 5**

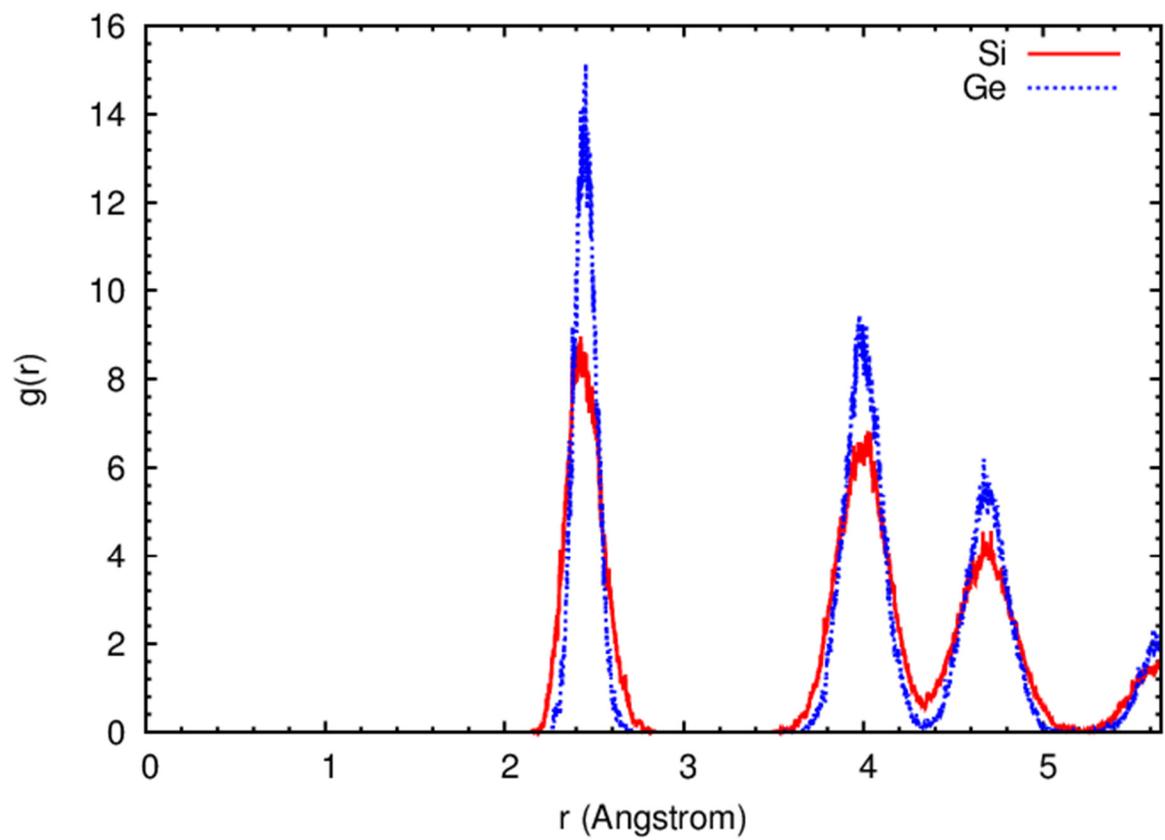

Fig. 6

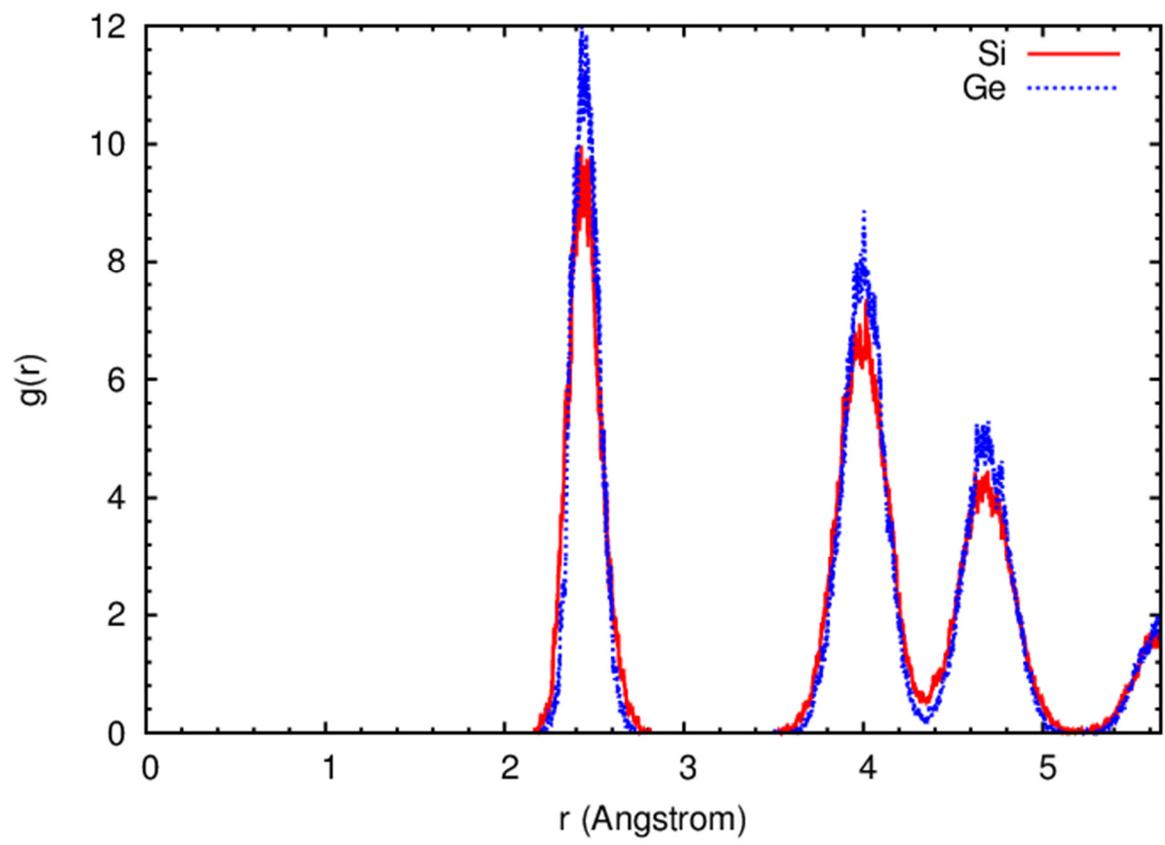

**Fig. 7**

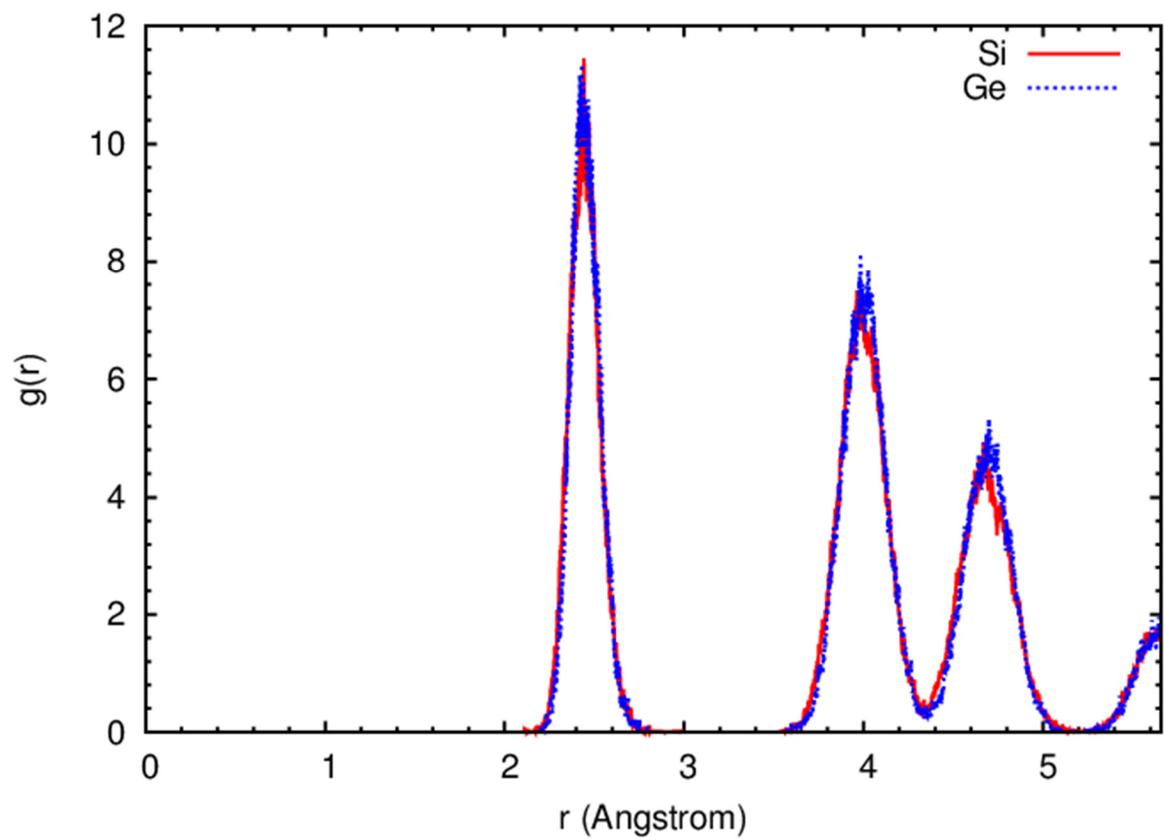

**Fig. 8**

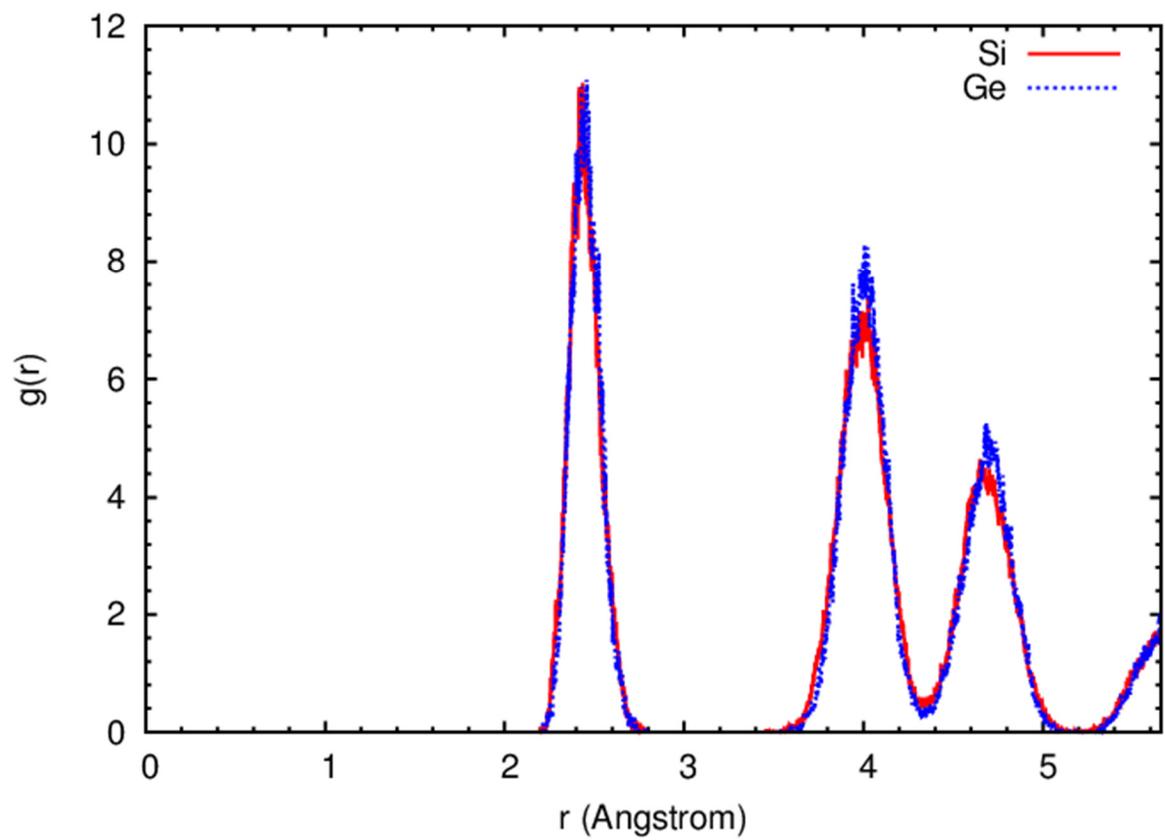

**Fig. 9**

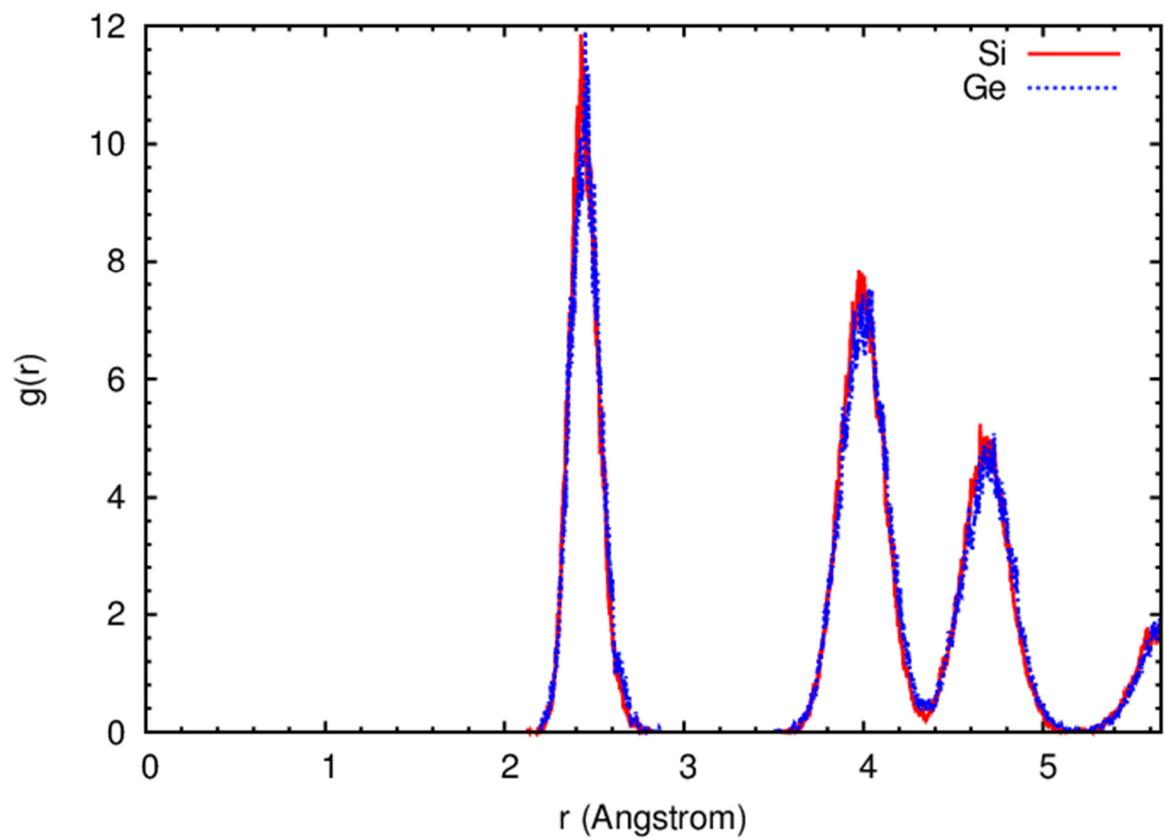

**Fig. 10**

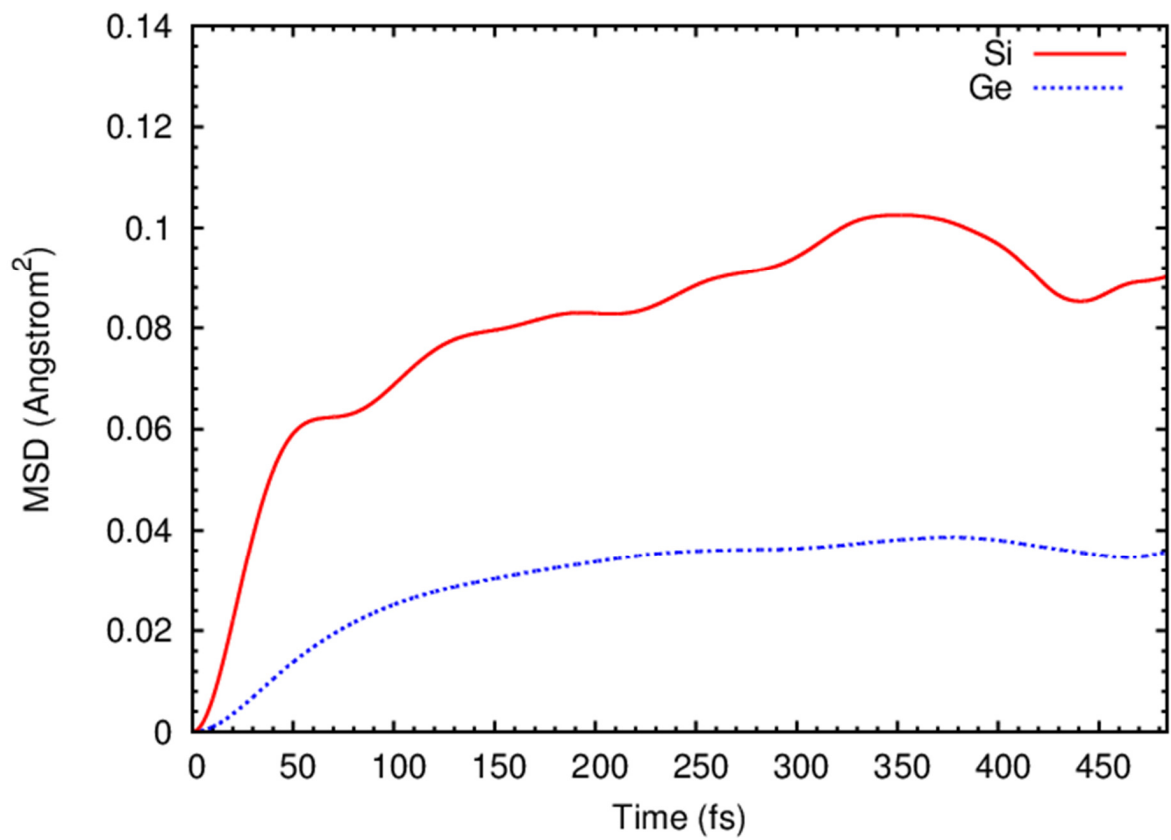

**Fig. 11**

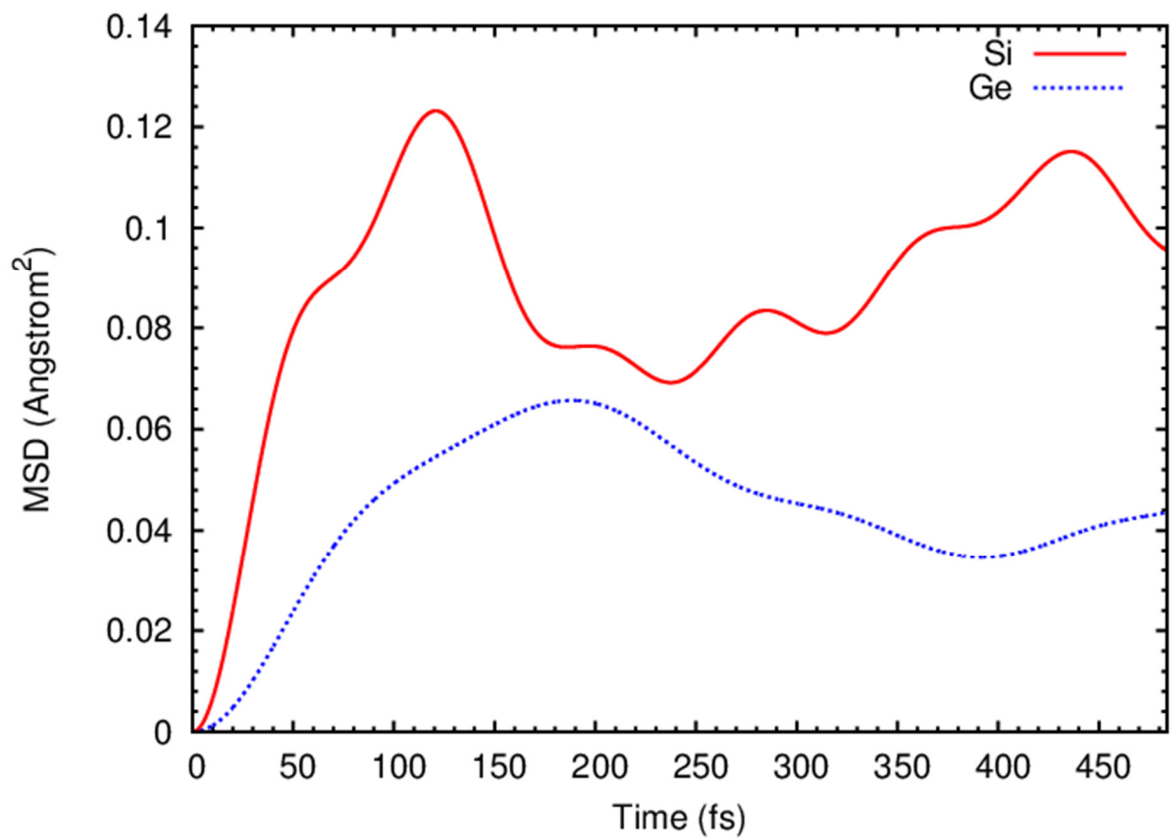

**Fig. 12**

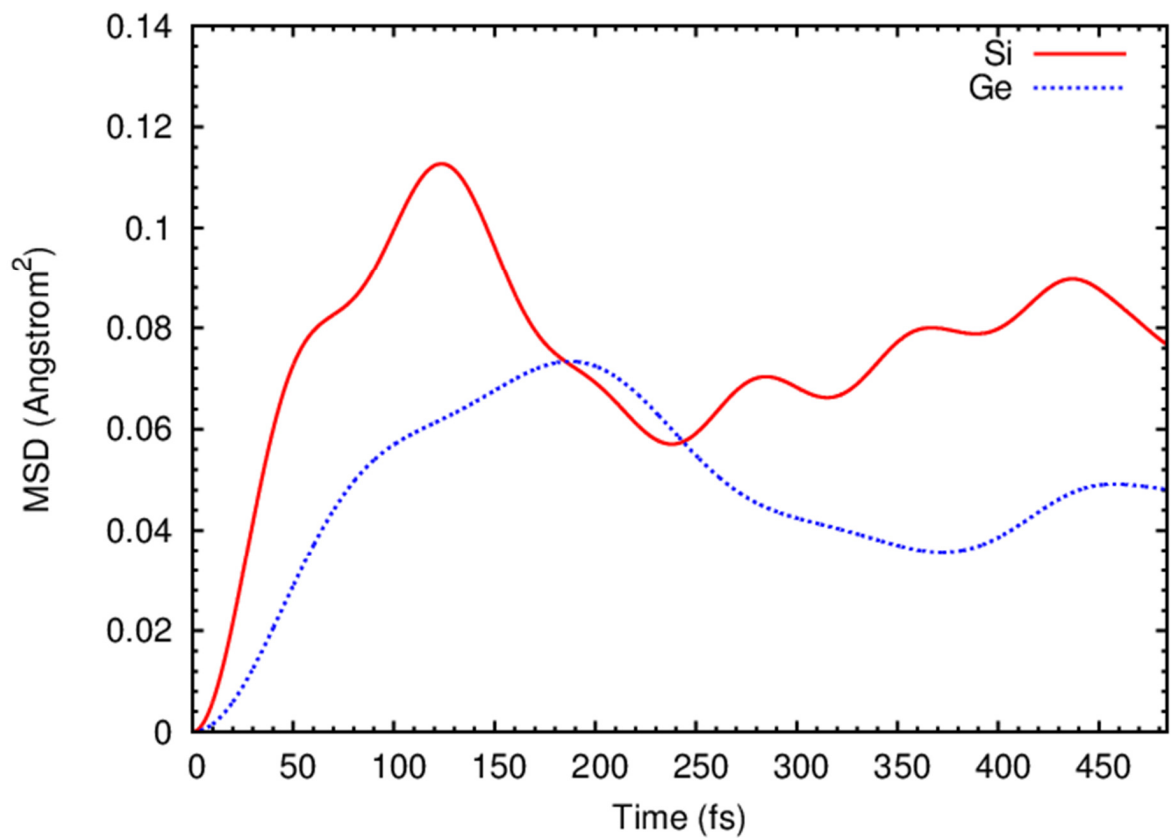

**Fig. 13**

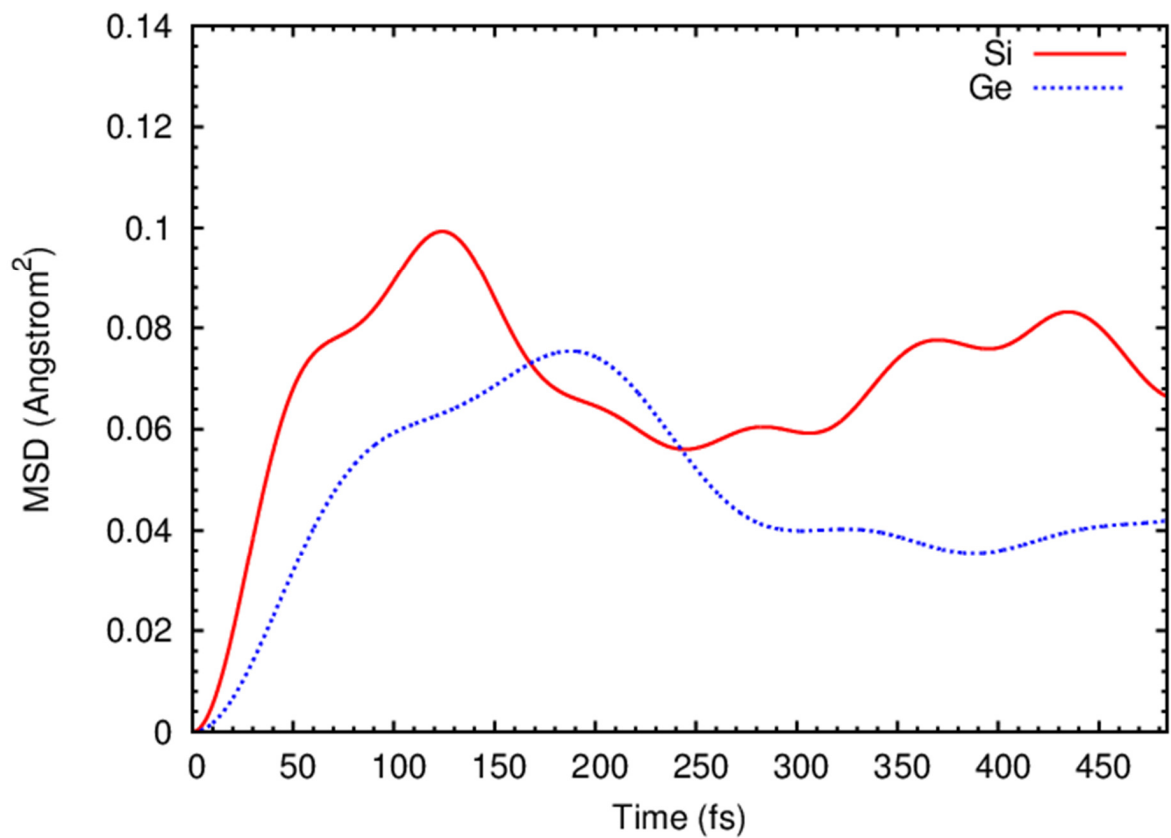

**Fig. 14**

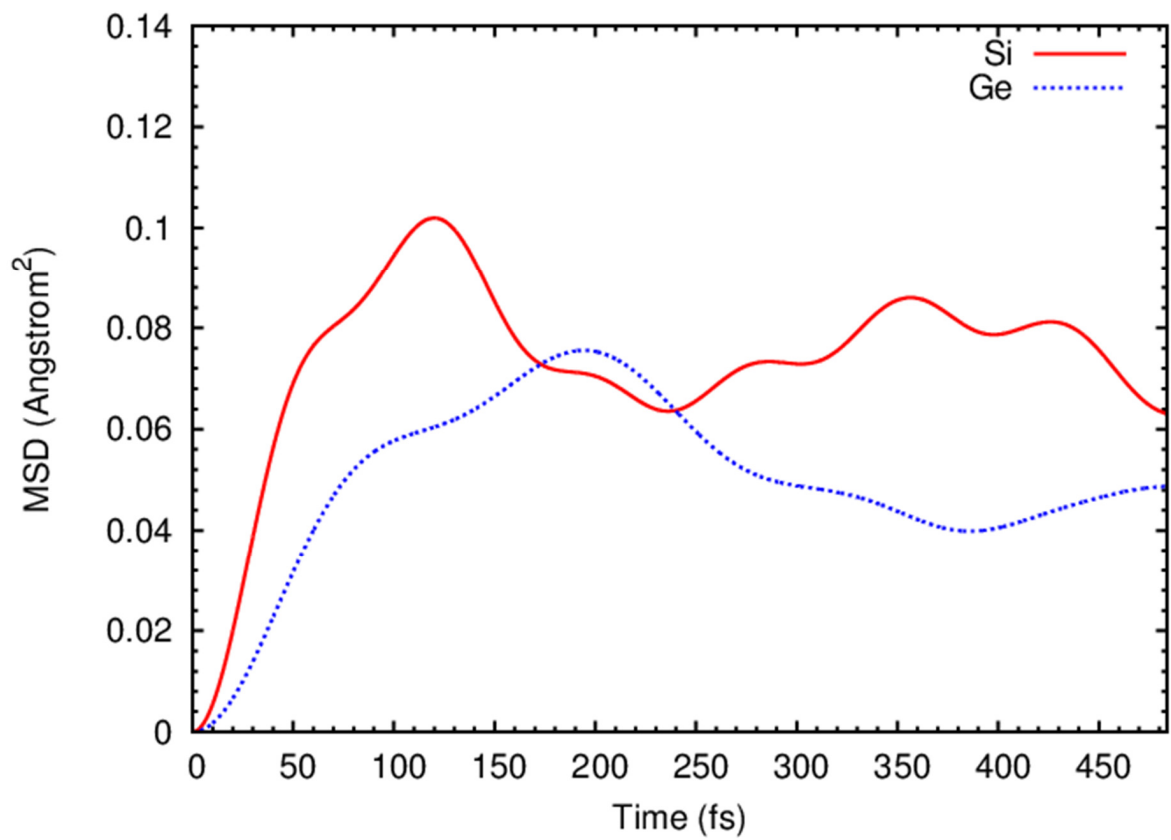

**Fig. 15**

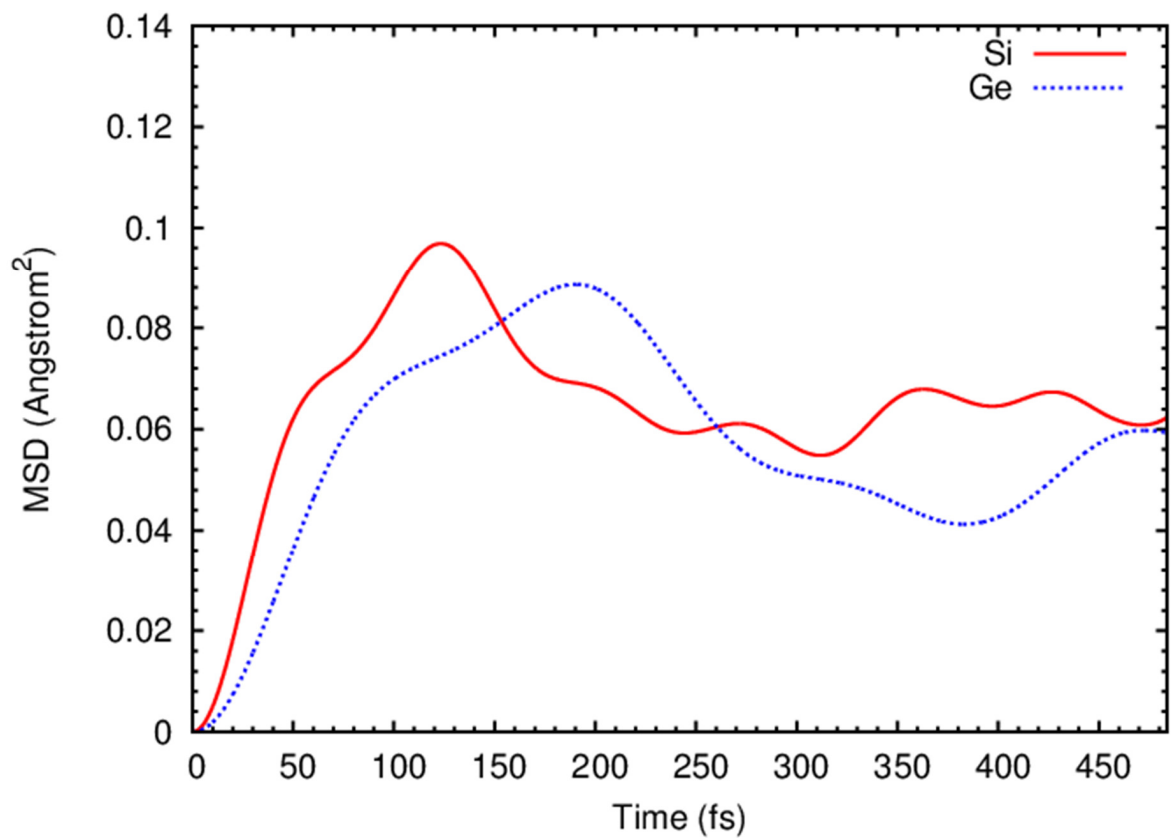

**Fig. 16**

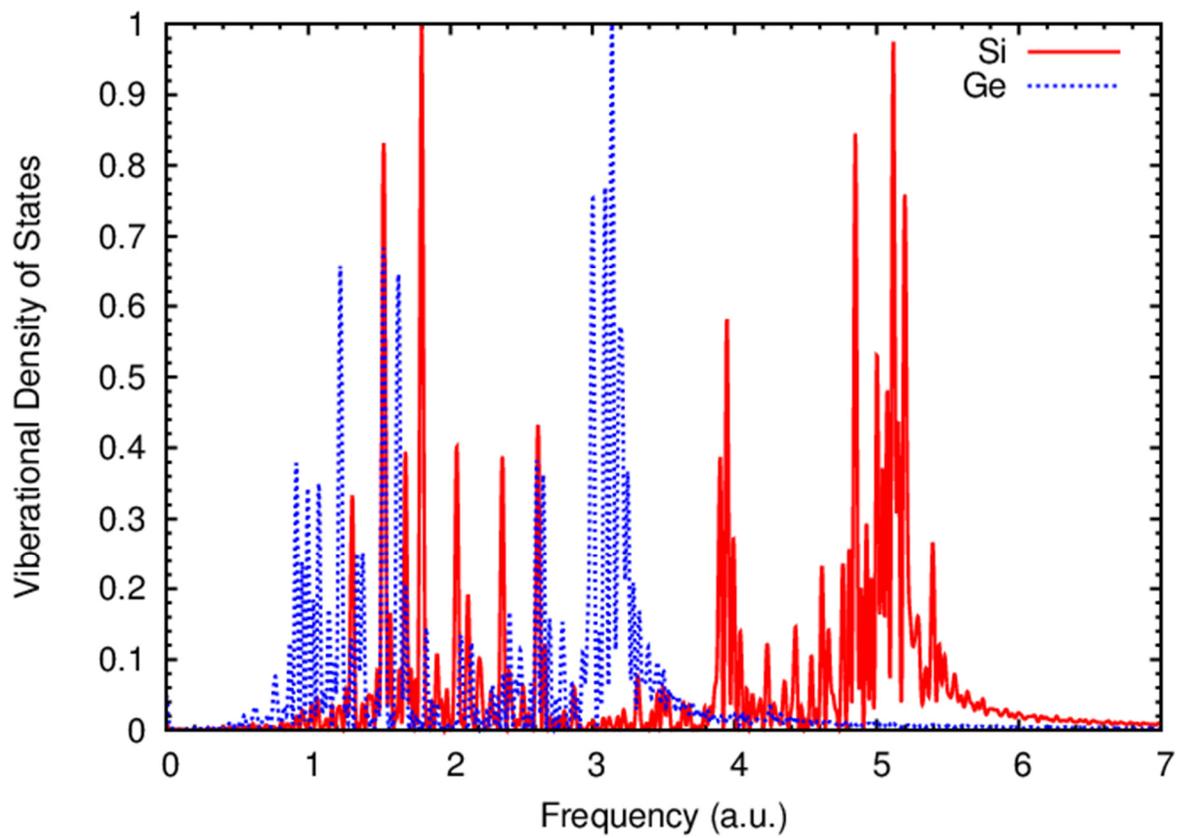

**Fig. 17**